\documentclass[headsepline]{scrartcl}

\usepackage[latin1]{inputenc}   
\usepackage{latexsym}
\usepackage{amssymb}
\usepackage{amsfonts}
\usepackage{amsmath}
\usepackage[numbers,sort&compress]{natbib}
\usepackage{bm}
\usepackage{graphicx}
\usepackage{hyperref}
\usepackage{color}
\usepackage{multicol}

 \providecommand{\dprod}{\!\cdot\!}%
 \providecommand{\wprod}{\!\wedge \!}
 
 \providecommand{\sfrac}[2]{\frac{#1}{#2}\,}
 \providecommand{\email}[1]{email: \href{mailto:#1}{\texttt{#1}}}

\numberwithin{equation}{section}

\begin{document}

\newcommand{\mytitle}{How much in the Universe can be explained by geometry?}
\title{\mytitle}

\author{José B. Almeida\\ Universidade do Minho, Physics
Department\\
Braga, Portugal, \email{bda@fisica.uminho.pt}}

\date{}

\pagestyle{myheadings} \markright{José B. Almeida \hfill \mytitle}


\maketitle

\begin{abstract}
The paper uses geometrical arguments to derive equations with
relevance for cosmology; 5-dimensional spacetime is assumed because
it has been shown in other works to provide a setting for
significant unification of different areas of physics. Monogenic
functions, which zero the vector derivative are shown to effectively
model electrodynamics and relativistic dynamics if one allows for
space curvature. Applying monogenic functions to flat space, the
Hubble relation can be derived straightforwardly as a purely
geometrical effect. Consideration of space curvature induced by mass
density allows the derivation of flat rotation curves for galaxies
without appealing for dark matter. Similarly, a small overall mass
density in the Universe is shown to provide a possible explanation
for recent supernovae observations, without the need for a
cosmological constant.
\end{abstract}


\section{Introduction}
Modern science is based on theories, each with its own application
domain; whenever we need to predict the outcome of an experiment or
observation we must resort to the relevant theory, insert the data
and derive the predictions which must then be in accord with the
experimental or observational results. A modern theory is based on a
set of principles or axioms, from which predictions are derived by
means of universally accepted rules of logic and/or mathematics. For
situations well within the boundaries of one theory's application
domain the system works well but problems usually arise when the
subject under scrutiny lies somewhere near the border between
theories. Furthermore, we tend to think that the Universe and
everything in it must somehow be ruled by global rules, applicable
everywhere at all times and this contradicts the co-existence of
separate theories' domains.

One common misconception is that one global theory for the Universe,
the so called theory of everything, would spell the end of science
for, by mastering such theory, one would in principle be able to
predict the outcome of any conceivable experiment or observation.
Things will probably never confirm this view, since science has many
overlapping layers such that high-level layers are virtually
independent from the way these layers are rooted on the underlying
ones. It is possible to make an analogy with computer science with
its programming languages in different levels. One can fully predict
the behaviour of a word processing application without ever
considering how bits are manipulated at machine level. There is even
a more profound meaning in this analogy, resulting from the fact
that we are able to identify a particular application implemented on
different platforms. The application has an identity which has
become detached from the platform and we have every reason to
believe a similar thing happens in our Universe, with some sciences
having become detached from the underlying structure.

In this paper we will address questions of cosmology in terms that
may allow a high degree of unification with other areas of physics,
namely with quantum mechanics. In physics one is usually confronted
with the need to opt between general relativity or quantum
mechanics, one theory being applicable to the very large and the
other one to the very small; both theories are expected to default
to Newtonian mechanics at the human scale. The situation is
uncomfortable from an intellectual standpoint but it is also rather
embarrassing when one tries to model the primordial Universe. At the
very early stages the Universe was very small and very dense; both
quantum mechanics and general relativity break down for such an
extreme situation and we must humbly recognize that a different and
yet unknown physics must apply. At present times, phenomena such as
are observed in active galaxy nuclei (AGN) also point to the need of
some sort of reconciliation between the two reigning theories of
physics.

\section{Physics from geometry}
Is physics ultimately the way we perceive a privileged geometry?
This is a fundamental question whose final answer is unknown but in
the affirmative case, if one day we are fortunate enough to find
that geometry, all the foundational equations of physics will
emanate from the symmetries, topology and other geometric
peculiarities. In spite of the fact that we can have an immediate
perception of 3-dimensional geometry, geometry at large is an
abstract mathematical subject which can only be expected to produce
equations of physical significance if a wise assignment is made
between geometrical and physical entities, namely the assignment
between geometric coordinates and the dimensions of physical space
and time. It is important to stress that we must be wise in the
assignment process, for an improper choice of coordinates may
completely hide from us the equations we would like to derive. As an
example, imagine we would like to study atmospheric currents but
were unaware of the Earth's spherical shape and rotation.
Accordingly we would probably choose a fixed Cartesian frame and in
no way would we be able to predict winds. A similar example is that
of Ptolomey's epicycloids to model planet's orbits; the solar
centered reference system revealed a much simpler picture than the
Earth centered counterpart.

The search for a putative privileged geometry of physics cannot be
done blindly amid the infinite variety of geometries at our
disposal; we are to be guided in this process by the work of others
and by our own previous efforts and intuition. Many hints point
towards the usefulness of exploring curved 5-dimensional spacetime
geometries, starting with the work by Kaluza \cite{Kaluza21} that
showed the way to the unification of general relativity with
electromagnetism. Kaluza's work was later complemented by Klein with
the postulate that one of the dimensions was to be taken as compact
but we will not be following that route. The author has also
proposed a unified approach to physics based on 5-dimensional
spacetime; a short introductory text is \citet{Almeida05:4};
applications to quantum mechanics can be found in
\citet{Almeida06:4,Almeida06:1} and an application to cosmology
anticipating the present work was published in 2006
\cite{Almeida05:3}.

The point of departure for our work is the exploitation of
\emph{monogenic functions} in 5-dimensional spacetime with signature
$(-++++)$. Monogenic functions zero the vector derivative of
geometric calculus\footnote{Detailed explanation of geometric
algebra and geometric calculus can be found in \citet{Doran03}; we
will need here only a small set of the mathematical tools offered by
this discipline and we will try to give the reader enough
information so that he does not need a profound knowledge of the
subject.} and as such can be seen as foundational. In the simplest
case of 1-dimensional space a monogenic function is a function with
zero derivative, that is, a constant with no structure; in
2-dimensional space monogenic functions are equivalent to analytic
functions and increasing the space dimensionality one finds that
those functions can generate many kinds of interesting geometrical
structures. To start with we assume there is an associative vector
product \emph{(geometric or Clifford product)} defined in this
space, with a commuting part and an anti-commuting part, such that
\begin{equation}
    a b = a \dprod b + a \wprod b, ~~~~ ba = a \dprod b - a \wprod
    b.
\end{equation}
The commuting part is a scalar called the \emph{inner product} and
the anti-commuting part is an oriented area called the \emph{outer
product}. We assume also a set of five independent vectors
$\{g_\alpha \}$, with $\alpha = 0, \ldots , 4$, used as a reference
frame. The metric tensor can then be constructed by
\begin{equation}
    g_{\alpha \beta} = g_\alpha \dprod g_\beta.
\end{equation}
The frame vectors are point dependent implying a point dependence
for the metric tensor; the latter has $25$ degrees of freedom, the
number required for modelling general relativity and
electromagnetism. The particular case of flat 5-dimensional
spacetime is characterized by having a metric tensor $g_{\alpha
\beta} = \mathrm{diag}(-1,1,1,1,1)$. The exterior product of the 5
frame vectors produces an imaginary quantity called a pseudo-scalar
\begin{equation}
    g_1 \wprod g_2 \wprod g_3 \wprod g_4 \wprod g_5 = \mathrm{i} v,
\end{equation}
with $\mathrm{i}^2 = -1$ and $v$ real.

The reference frame is associated with a \emph{reciprocal frame},
denoted with upper indices, whose vectors verify the relation
\begin{equation}
    g^\alpha \dprod g_\beta = \delta^\alpha _\beta,
\end{equation}
where $\delta^\alpha _\beta$ is the Kronecker delta. The inner
product of the reciprocal frame vectors produces a reciprocal metric
tensor
\begin{equation}
    g^\alpha \dprod g^\beta = g^{\alpha \beta} = (g_{\alpha
    \beta})^{-1}.
\end{equation}
We use the reciprocal frame for the definition of a derivative
operator called the \emph{vector derivative}; the definition is
\begin{equation}
\label{eq:vector_deriv}
    \mathrm{D} = g^\alpha \partial_\alpha,
\end{equation}
where we made use of the summation convention and the compact
notation for partial derivatives $\partial_\alpha =
\partial/\partial_\alpha$. The vector derivative can be applied to
all geometric entities, in particular to scalars and vectors as
follows:
\begin{itemize}
    \item divergence of a scalar function, $\mathrm{D}\psi$,

    \item gradient of a vector function, $\mathrm{D} \dprod \psi$,

    \item exterior derivative of a vector function, $\mathrm{D}
    \wprod \psi$.
\end{itemize}
The exterior derivative is an oriented area which in the special
case of 3-dimensional Euclidean space can be represented by the
length of a vector normal to it; this is why in this space we can
use the concept of curl as a the result of applying a differential
operator to a vector.

The vector derivative allows us to define a special class of
functions, called \emph{monogenic functions}: those that have a null
derivative. A monogenic function verifies the equation
\begin{equation}
\label{eq:monogenic}
    \mathrm{D} \psi = 0.
\end{equation}
There is a large variety of solutions to this equation, which we
cannot explore in this work. In \citet{Doran03} the solutions for
3-dimensional Euclidean flat space are shown to produce spherical
harmonics and in \cite{Almeida06:4,Almeida06:1} special solutions
for 5-dimensional spacetime are used for the derivation of quantum
mechanics' equations. In this paper we will be particularly
interested in the ability of monogenic functions to model dynamics;
the next section is dedicated to this subject.

\section{Dynamics by monogenic functions}
If the vector derivative is multiplied by itself the result is a
second order scalar operator; we know that it is a scalar because it
results from the square of a vector. A scalar operator does not
alter the geometric character of the function to which it is
applied; we call this second order operator a Laplacian and its
definition is quite simply
\begin{equation}
    \mathrm{D}^2 \psi = \mathrm{D}(\mathrm{D}\psi).
\end{equation}
Using definition (\ref{eq:vector_deriv}) we can write
\begin{equation}
    \mathrm{D}^2 \psi = g^\alpha \partial_\alpha (g^\beta
    \partial_\beta \psi).
\end{equation}
In many practical situations the frame vectors will be slowly
varying functions by comparison to function $\psi$ and when this
happens we are allowed to neglect the former's derivatives
($\partial_\alpha g^\beta \approx 0$). The Laplacian then simplifies
to
\begin{equation}
\label{eq:d2_psi}
    \mathrm{D}^2 \psi = g^{\alpha \beta} \partial_{\alpha \beta}
    \psi.
\end{equation}

Any monogenic function is by necessity a null Laplacian function;
this can be verified easily by dotting Eq.\ (\ref{eq:monogenic})
with $\mathrm{D}$. On the other hand a function can have null
Laplacian without being monogenic; however the second order equation
resulting from a null Laplacian is usually easier to solve than its
first order monogenic counterpart; in many cases one can start by
imposing a null Laplacian complementing that with a verification of
the further restrictions imposed on the solutions by the monogenic
condition. When studying dynamics this approach is usually
recommended, so let us examine solutions for Eq.\ (\ref{eq:d2_psi}).
We are immediately led to try a solution of the type
\begin{equation}
\label{eq:psi}
    \psi = \psi_0 \mathrm{e}^{\mathrm{i} p \cdot x},
\end{equation}
with $p = g^\alpha p_\alpha$ and $x = g_\beta x^\beta$. Inserting
into Eq.\ (\ref{eq:d2_psi}) we see that
\begin{equation}
\label{eq:null_p}
    g^{\alpha \beta} p_\alpha p_\beta = p^2 = 0,
\end{equation}
as long as the derivatives of $p$ can be neglected with respect to
the derivatives of $\psi$. Equation (\ref{eq:psi}) includes a
constant factor $\psi_0$ which can be any geometric entity if only
the null Laplacian condition is imposed; it is the monogenic
condition that clarifies the geometric character of $\psi_0$. Being
concerned only with dynamics we can ignore this issue which becomes
relevant in quantum mechanics.

Equation (\ref{eq:null_p}) tells us that vector $p$ is null but we
need to assign physical significance to its components in order to
interpret this fact physically. We propose a decomposition of vector
$p$ as
\begin{equation}
    p = g^0 E + \mathbf{p} + g^4 m,
\end{equation}
where $E$ is total energy, $\mathbf{p}= g^1 p_1 + g^2 p_2 + g^3 p_3$
is 3-dimensional momentum and $m$ is rest mass. It is particularly
interesting to examine the case of flat spacetime, when the null
condition for $p$ produces
\begin{equation}
    E^2 = \mathbf{p}^2 + m^2,
\end{equation}
a well known relation from special relativity.

If we were working in the 4-dimensional spacetime of general
relativity, a function $\psi$ given by Eq.\ (\ref{eq:psi}) would be
interpreted as an electromagnetic wave and the metric tensor would
predict light bending by gravity but here we have one extra
dimension, so some work must be done in order to arrive at a
physical interpretation. In order to understand how the null
Laplacian generates dynamics we search for the conditions that must
be met for $\psi$ to remain constant. Obviously this implies that
the inner product $p \dprod x$ in the exponent remains constant or,
equivalently $\mathrm{d}(p \dprod x) = 0$. Since we are only
considering situations where $p$ remains essentially invariant, we
can write for constant $\psi$
\begin{equation}
    p \dprod \mathrm{d}x = 0.
\end{equation}
The left hand side of this equation is an inner product of two
vectors, which can be null in two circumstances: either the two
vectors are perpendicular to each other or they are both null. If
$\mathrm{d}x$ is perpendicular to $p$ we are defining a wavefront,
which here has 3 dimensions and can be called an hyperplane;
accordingly $\psi$ is said to be a quasi-hyperplane wave. Since it
has been established by Eq.\ (\ref{eq:null_p}) that $p$ is null, the
other possibility for constant $\psi$ happens when $\mathrm{d}x$ is
also null, that is when
\begin{equation}
\label{eq:null_dx}
    g_{\alpha \beta} \mathrm{d}x^\alpha \mathrm{d}x^\beta = 0.
\end{equation}
This is all we need in order to express dynamics but we can make it
more familiar if we assume that $g_{4 \mu} = 0$, $\mu = 0, \ldots ,
3$. The equation can then be rewritten as
\begin{equation}
\label{eq:g_relativity}
    (\mathrm{d}x^4)^2 = \sfrac{g_{\mu \nu}}{g_{44}} \mathrm{d}x^\mu
    \mathrm{d}x^\nu, ~~~~ \mu, \nu = 0, \ldots ,3.
\end{equation}
The reader will recognize in this equation the quadratic form of
general relativity with a metric specified by $g_{\mu \nu}/g_{44}$;
this will only work if the frame vectors $g_\alpha$ are independent
from coordinate $x^4$. A Lagrangian can be associated with this
quadratic form and geodesic trajectories derived from it so,
naturally, we have a means of modelling gravitational dynamics. For
this purpose it is legitimate to associate coordinate $x^0$ to time
and coordinate $x^4$ to proper time; note however that a similar
association may work only for gravitational dynamics an not for
electrodynamics or quantum mechanics. When convenient we will denote
coordinate $x^0$ with the letter $t$ and coordinate $x^4$ with the
letter $\tau$.

Equation (\ref{eq:null_dx}) is a scalar equation which we can
manipulate in a different way to what was done above. A useful
alternative consists consist in isolating $(\mathrm{d}x^0)^2$ in the
left hand side to obtain
\begin{equation}
\label{eq:4do}
    (\mathrm{d}x^0)^2 = \sfrac{g_{i j}}{g_{00}} \mathrm{d} x^i
    \mathrm{d}x^j, ~~~~ (i, j) = 1, \ldots , 4,
\end{equation}
where we have assumed that $g_{0i} = 0$ and the frame vectors are
independent from $x^0$. We have now obtained the quadratic form of
4-space with Euclidean signature and an equation similar to the
eikonal equation we are used to seeing in geometrical optics but
with 4 dimensions instead of the usual 3. We call this
\emph{4-dimensional optics} for obvious reasons. There are cases
when we can obtain both quadratic forms and in those cases we have
the choice of performing an analysis in terms of general relativity
or 4-dimensional optics; the results are coincident but the
perspective is entirely different. This is the case for all static
metrics in general relativity; consequently they can also be
examined under a 4-dimensional optics approach. In physics, looking
at a problem with different approaches usually enlarges our
understanding about that problem; a 4-dimensional optics approach
becomes especially revealing in quantum mechanics, because
quantization can be seen as akin to propagation modes in
4-dimensional waveguides.

\section{Hyperspherical symmetry}
Dynamics in curved space is governed by Eq.\ (\ref{eq:null_dx}), as
we have seen. However many important conclusions can be drawn from
the analysis of flat space; here we should talk about kinematics
rather than dynamics, because we will be dealing only with bodies in
inertial movement, modelled by hyperplane 4-dimensional waves.
Obviously all hyperplane waves propagate along straight lines and
there is apparently not much to say about them; we shall see that
things are not as simple as they seem and a few surprises will
emerge from this analysis.

Returning to Eq.\ (\ref{eq:null_dx}) but rewriting for flat space we
get
\begin{equation}
\label{eq:light_cone}
    -(\mathrm{d}x^0)^2 + \sum_{i=1}^4 (\mathrm{d}x^i)^2 =0,
\end{equation}
which is a specification for null displacements or displacements on
the light hypercone. This is easier to understand if we downgrade to
3-dimensional spacetime; in 3 dimensions geometrical representation
becomes possible, so it is usually useful to start with
3-dimensional analysis and progress gradually to higher dimensions.
A null displacement in 3-dimensional spacetime is specified by
$(\mathrm{d}x^0)^2 - (\mathrm{d}x^1)^2 -(\mathrm{d}x^2)^2 = 0$. All
null displacements take place on the surface of a cone with apex at
the current position and this cone is usually called the light cone,
because light travels along null displacement paths. If we want to
restrict our attention to null displacements we don't need 3
coordinates because we know beforehand that we are restricted to a
2-dimensional surface. Since the cone's axis is parallel to $x^0$ we
can specify any displacement on its surface by a radial displacement
(distance to the axis) and a polar angle variation. We have just
switched to cylindrical polar coordinates because they are the most
convenient to express the symmetry implied by null displacements. If
we denote the radial and angular coordinates centered at the current
position by $\tau$ and $\rho$, respectively, the null displacement
condition becomes $(\mathrm{d}x^0)^2 - (\mathrm{d}\tau)^2  - \tau^2
(\mathrm{d}\rho)^2= 0$.

Physical spacetime is 4-dimensional and so null displacements are
performed on a 3-dimensional surface to which we can call a light
hypercone; in special relativity one usually refers to the light
cone, in spite of the fact that this is a 3-dimensional
hypersurface, but in this paper it is convenient to make the
distinction clear. Again if we are interested only in null
displacements 4 coordinates are one too many and the equations will
gain clarity if we adopt spherical polar coordinates. One dimension
above, the situation specified by Eq.\ (\ref{eq:light_cone}) is
again one of null displacements but now in 5-dimensional spacetime.
One sees immediately that all displacements are now on the surface
of a 4-dimensional light hypercone and that the appropriate
coordinates to bring out the implied symmetry are hyperspherical
polar coordinates, comprising one radial distance and 3 polar
angles; we denote those coordinates by $\{\tau, \rho, \theta,
\phi\}$, respectively. In hyperspherical coordinates a null
displacement is expressed by
\begin{equation}
\label{eq:null_hspherical}
    -(\mathrm{d}t)^2 + (\mathrm{d}\tau)^2 + \tau^2
    [(\mathrm{d}\rho)^2 + \sin^2 \rho (\mathrm{d}\theta)^2 + \sin^2
    \rho \sin^2 \theta (\mathrm{d}\phi)^2] = 0.
\end{equation}
Here we see that  the association of coordinate $\tau$ with proper
time implies that the 3 coordinates associated with physical space
are turned into distances measured on the surface of a 3-sphere so,
when we use Cartesian coordinates for physical space we are actually
referring to points on the hyperplane tangent to the hypersphere and
not to points of constant $\tau$. We can make an analogy to
displacements on the Earth's surface, which can be approximated with
displacements on a plane locally tangent to Earth, as long as
displacements remain small. What we are saying here is that
Cartesian coordinates can only be used on a local scale but not for
cosmological distances.

\section{\label{fsd}Flat space kinematics}
One case of special interest considers displacements with a common
origin. If displacements start at the apex of the light hypercone,
they can only follow generatrices characterized by $\mathrm{d}\rho =
\mathrm{d}\theta = \mathrm{d}\phi = 0$ and have a very simple null
condition given by
\begin{equation}
    (\mathrm{d}t)^2 = (\mathrm{d}\tau)^2.
\end{equation}
For convenience we will set $\tau = t$, that is, the coordinate's
origin is placed at the hypercone's apex. Inserting this into the
equation above and denoting derivatives with respect to $t$ by a
dot, we have
\begin{equation}
    \dot{\tau} = 1,
\end{equation}
for displacements along a generatrix.

All generatrices diverge from the hypercone's apex; naturally the
distance between two generatrices increases with the radial
coordinate $\tau$. If we wish to compute distances to a particular
point, measured over the hyperspherical directrix,  we can set the
origin for coordinate $\rho$ at that point and introduce a distance
coordinate
\begin{equation}
    r = \tau \rho.
\end{equation}
A simple manipulation allows us to write
\begin{equation}
    \dot{r} = \dot{\tau} \rho + \tau \dot{\rho} = \dot{\tau} \rho =
    \sfrac{\dot{\tau}}{\tau} r.
\end{equation}
But we have seen above that $\dot{\tau}$ is unity, so the previous
equation states that two generatrices are coming apart with a
velocity $\dot{r}$ proportional to their distance $r$; this is
exactly what the Hubble relation says about galaxies and we are led
to define the Hubble parameter by
\begin{equation}
    H = \sfrac{1}{\tau}.
\end{equation}
This is an extremely important conclusion. A physical interpretation
of our geometric argument allows us to say that the wavefunctions of
cosmical objects, such as galaxies or even galaxy clusters, are
monogenic functions of 5-dimensional spacetime, diverging from a
common origin, the big bang. As a consequence the Universe has an
overall hyperspherical symmetry and we are wrong when we choose
Cartesian coordinates for its description. By choosing the
appropriate hyperspherical polar coordinates we detect immediately
that distances between cosmical objects must grow at a rate which is
exactly proportional to those distances. The Hubble relation thus
acquires a purely geometrical explanation and results from the
consideration of an empty Universe; there is no need for a critical
mass density to justify this flat rate expansion.

In this model cosmological objects such as galaxy clusters are still
objects, evolving along generatrices of the light hypercone. This
contradicts the equivalence principle, in the sense that we now have
an absolute definition of motion; our privileged frame, the frame of
absolute stillness, is provided by all the galaxy clusters in spite
of their apparent relative motion. Still objects are characterized
by the relation $\dot{\tau} = 1$, while objects in motion relative
to the privileged frame are characterized by $\dot{\tau} < 1$. From
Eq.\ (\ref{eq:null_hspherical})we derive
\begin{equation}
    1 - \tau^2 [\dot{\rho}^2 + \sin^2 (\rho) \dot{\theta}^2
     + \sin^2 (\rho) \sin^2 (\theta) \dot{\phi}^2 ]- \dot{\tau}^2 = 0.
\end{equation}
A physical interpretation of this equation becomes easier if we make
the replacement $\tau \rho \rightarrow r$:
\begin{equation}
\label{eq:velocity_general}
    1 - \left[1 + \left(\sfrac{r}{\tau} \right)^2
    \right]\dot{\tau}^2 -\dot{r}^2  + 2 \sfrac{r}{\tau} \dot{\tau}\dot{r} -
    r^2[\dot{\theta}^2 + \sin^2(\theta) \dot{\phi}^2] = 0.
\end{equation}
What we usually designate by velocity is the quantity $v$ verifying
the relation
\begin{equation}
    v^2 = \dot{r}^2  +  r^2[\dot{\theta}^2 + \sin^2(\theta)
    \dot{\phi}^2].
\end{equation}
Replacing above we get
\begin{equation}
\label{eq:compatibility}
    1 - v^2 - \left[1 + \left(\sfrac{r}{\tau} \right)^2
    \right]\dot{\tau}^2  + 2 \sfrac{r}{\tau} \dot{\tau}\dot{r} = 0.
\end{equation}
In laboratory experiments $r \ll \tau$, the equation simplifies to
$\dot{\tau}^2 + v^2 = 1$ and we are taken back to special
relativity.

Equation (\ref{eq:compatibility}) expresses the compatibility
between relativistic physics, which applies in laboratory
experiments including those of high energy physics, and the physics
of cosmology, which we must start to consider when the distances to
our observational point become comparable to the size of the
Universe. The orbits of planets involve distances which are still
small compared to the Universe and we expect relativistic dynamics
to work right for them but what about the dynamics of galaxies? We
shall have a brief look at this subject in the next section.

\section{Rotation curves of galaxies}
The subject of galaxies' dynamics is a very complex one and we will
not dare to examine it in any depth in this paper. In this section
we will only have a very superficial look at some possible
consequences that arise from our approach to dynamics in general via
monogenic functions and 5-dimensional spacetime. First of all we
must check that our approach is compatible with what is generally
known about orbits under gravitational field; in particular we must
show that the angular velocity of a circular orbiting body varies
with $r^{-3/2}$, $r$ being the orbital radius.

Equation (\ref{eq:g_relativity}) allows us to introduce a metric
such as Schwarzschild's in order to model the gravitational
interaction but we will instead Yilmaz's metric \cite{Yilmaz58,
Yilmaz71}, since it mathematically more convenient and it is
equivalent to Schwarzschild's for the sort of distances we will be
dealing with. In spherical coordinates the quadratic form of
Yilmaz's metric can be written as
\begin{equation}
    (\mathrm{d}\tau)^2 = \mathrm{e}^{-2m/r} (\mathrm{d}t)^2 +
    \mathrm{e}^{2m/r}[(\mathrm{d} r)^2 + r^2 (\mathrm{d}\theta)^2 +
    r^2 \sin^2 \theta (\mathrm{d}\phi)^2],
\end{equation}
where $m$ is a spherical mass, $r$ is the distance to the center of
gravity of mass $m$ and we are using Planck or non-dimensional units
\cite{Almeida05:3}. We could use the previous equation to derive
circular orbits but it will be more convenient for the development
of our argument to rewrite it in the form of Eq.\ (\ref{eq:4do}); it
is then
\begin{equation}
    (\mathrm{d}t)^2 = \mathrm{e}^{2m/r} (\mathrm{d}\tau)^2
    + \mathrm{e}^{4m/r}[(\mathrm{d} r)^2 + r^2 (\mathrm{d}\theta)^2 +
    r^2 \sin^2 \theta (\mathrm{d}\phi)^2].
\end{equation}
Orbital equations can now be found by searching for geodesics of
this space; these result from consideration of the Lagrangian
defined by \cite{Inverno96, Martin88}
\begin{equation}
   1 = 2 L = \mathrm{e}^{2m/r} \dot{\tau}^2 +
    \mathrm{e}^{4m/r} [\dot{r}^2 + r^2 (\dot{\theta}^2 + \sin^2
    \theta \dot{\phi}^2)];
\end{equation}
we suppressed the parenthesis around the coordinates because there
are no upper indices to cause confusion. This equation is usually
simplified with the consideration that orbits are flat, so we are
free to set $\theta = \pi/2$. Since we are interested in circular
orbits we can also set $\dot{r} = 0$; the simplified Lagrangian then
verifies
\begin{equation}
\label{eq:orbit}
   1 = 2 L = \mathrm{e}^{2m/r} \dot{\tau}^2 +
    \mathrm{e}^{4m/r}  r^2  \dot{\phi}^2.
\end{equation}
The generalized momentum associated with $r$ is constant, so the
corresponding Euler Lagrange equation is $\partial_r L = 0$, which
expands to
\begin{equation}
    -\sfrac{m}{r^3} \dot{\tau}^2 +\left(1- \sfrac{2m}{r}\right)
    \mathrm{e}^{2m/r}  \dot{\phi}^2 =
    0.
\end{equation}
Expanding the exponential in series, taking just two terms and
simplifying we get
\begin{equation}
    \dot{\phi}^2 = \sfrac{m}{r^3} \dot{\tau}^2,
\end{equation}
which is the expected relation from Newtonian dynamics with a
relativistic correction given by the $\dot{\tau}^2$ factor.

Although we have based our deductions on monogenic functions, the
results obtained are the same as could have been derived with
standard general relativity; we have not even considered the
hyperspherical symmetry that we expect to result from the imposition
of the monogenic condition in 5-dimensional spacetime. If a
gravitational system as a whole verifies a single monogenic
function, however, we expect it to evolve on the light hypercone
with apex at the center and at the time when the system was
originated; we can check how such a system should evolve and try to
derive some consequences for cosmological objects. Assuming the
Yilmaz metric remains valid, the null displacement condition
(\ref{eq:null_dx}) must be written as
\begin{equation}
    -\mathrm{e}^{-2m/r}(\mathrm{d}t)^2 + (\mathrm{d}\tau)^2 +
    \mathrm{e}^{-2m/r} \tau^2 [(\mathrm{d}\rho)^2 + \sin^2 \rho
    (\mathrm{d}\theta)^2 + \sin^2 \rho \sin^2 \theta
    (\mathrm{d}\phi)^2] = 0.
\end{equation}
The flattness of orbits allows us to set $\theta = \pi/2$, as
before. We also argue that angle $\rho$ is small, for which reason
its sine can be replaced by the argument. With these simplifications
we are led to consider the geodesic Lagrangian
\begin{equation}
    1 = 2 L = \mathrm{e}^{2m/r} \dot{\tau}^2 + \mathrm{e}^{4m/r}
    (\dot{\rho}^2 + \rho^2 \dot{\phi}^2).
\end{equation}
The right hand side now includes both $\rho$ and $r$ but this can be
avoided if we recall that $r = \tau \rho$; making the substitution
we get
\begin{equation}
\label{eq:orbit_2}
    1 = 2L = \mathrm{e}^{2m/r} \dot{\tau}^2 + \mathrm{e}^{4m/r}
    \left[\left(\dot{r} - \sfrac{r}{\tau} \dot{\tau} \right)^2 + r^2
    \dot{\phi}^2 \right].
\end{equation}
Notice that this equation is written for pseudo-Euclidean space with
signature $(++++)$ and does not have a general relativity
counterpart because the metric tensor is dependent on coordinate
$\tau$; therefore it does not verify the static metric condition
that we invoked in order to be able to isolate in the left hand side
either $(\mathrm{d}x^0)^2$ or $(\mathrm{d}x^4)^2$.

Comparing Eqs. (\ref{eq:orbit_2}) and (\ref{eq:orbit}), we see that
the difference lies in the extra term with the parenthesis $(\dot{r}
- r \dot{\tau}/\tau)$. This parenthesis is null when $\dot{r}/r =
\dot{\tau}/\tau$, that we recognize as the Hubble relation seen
above. When this happens the rotational velocity verifies the very
simple relation
\begin{equation}
    r^2 \dot{\phi}^2 = 1 - \mathrm{e}^{2m/r} \dot{\tau}^2.
\end{equation}
The left hand side is the linear rotational velocity while the right
hand side is nearly constant for large values of $r$; this is
exactly the behaviour of orbiting material in the periphery of most
galaxies. The standard explanation for this behaviour of galaxies'
rotational curves invokes large amounts of unexplained dark matter
but we see here that no dark matter at all needs to be invoked; it
only needs a small expansion of the orbit, at a rate similar to the
Hubble expansion. The calculations for the periphery of {M 31}
galaxy predict an expansion velocity of $2.43\text{ Km s}^{-1}$ in
order to obtain a flat rotation curve; this is to be compared with a
rotational velocity around $300\text{ Km s}^{-1}$ and is
undetectable by present day instruments. In summary, if expansion of
the order of $1\%$ of rotational velocity is detected in galaxies
with flat rotation curves, this may mean that such galaxies are
governed by a single monogenic function and no dark matter at all is
needed to account for the observed dynamics.

\section{Cosmic evolution}

In Sec.\ \ref{fsd} we verified that a monogenic function in flat
space justified the option for hyperspherical coordinates and these
in turn led us to the Hubble relation; this is compatible with a
model for an empty Universe governed by a single monogenic function.
If this works for an empty Universe, can we make it work for a
Universe with a small mass density, of the order of the mass that
can effectively be detected, and do away with dark matter and dark
energy? In order to answer this question we recall Eq.\
(\ref{eq:orbit_2}) and consider the case of pure expansion, so that
$\dot{\phi}=0$.

A model for the whole Universe can be constructed if we take a thin
spherical shell of mass $m_s$ and radius $r$ centered on any point
of the Universe. The gravitational pull on the spherical shell is
exerted by the total mass $m_i$ inside the shell, which is obviously
\begin{equation}
    m_i = \sfrac{4}{3} \pi r^3 \Omega,
\end{equation}
where $\Omega$ is the mass density. Equation (\ref{eq:orbit_2})
allows us to write the expansion rate for the shell as
\begin{equation}
\label{eq:expansion}
    \left(\sfrac{\dot{r}}{r}\right)^2  = \sfrac{\mathrm{e}^{-4m_i/r}
    -\mathrm{e}^{-2m_i/r}\dot{\tau}^2}{r^2}-
    \left(\sfrac{\dot{\tau}}{\tau}\right)^2
    + \sfrac{2 \dot{\tau}\dot{r}}{ \tau r}.
\end{equation}
We need to check that this equation is compatible with the Hubble
relation found earlier. In an empty Universe $m_i = 0$ and
$\dot{\tau}=1$; the equation simplifies to $\dot{r}/r = 1/\tau$, as
expected. In the standard cosmological model the flat expansion of
the Universe is attributed to a critical mass density $\Omega_c$
verifying
\begin{equation}
\label{eq:friedman}
        \left(\sfrac{\dot{r}}{r}\right)^2 = \sfrac{8 \pi \Omega_c}{3}.
\end{equation}
In our hyperspherical model the critical mass density is not mass at
all, but the two approaches can be made compatible if we define
\begin{equation}
\label{eq:critical}
        \Omega_c = \sfrac{3}{8 \pi \tau^2}.
\end{equation}

In the presence of a small mass density we have a perturbation of
the expansion rate caused by the first term on the right hand side
of Eq.\ (\ref{eq:expansion}). The exponents in this term include the
factor $m_i/r = 4 \pi r^2 \Omega/3$; inserting the definition
(\ref{eq:critical}) we get
\begin{equation}
    \sfrac{m_i}{r} = \sfrac{\mu \Omega_c r^2}{2 \tau^2},
\end{equation}
with $\mu = \Omega/\Omega_c$. Inserting into Eq.\
(\ref{eq:expansion}) and taking only the first order terms for the
exponentials, the perturbation term becomes
\begin{equation}
    -\mu \Omega_c  \left(\sfrac{\dot{\tau}}{\tau}\right)^2.
\end{equation}
Taking as an approximation that the two last terms on the right hand
side of Eq.\ (\ref{eq:expansion}) are still given by Eq.\
(\ref{eq:friedman}) we get
\begin{equation}
    \left(\sfrac{\dot{r}}{r} \right)^2 \approx \left[\sfrac{8 \pi}{3} - \mu
    \left(\sfrac{\dot{\tau}}{\tau}\right)^2 \right] \Omega_c.
\end{equation}
The important thing to note here is that the expansion rate
$\dot{r}/r$ is dependent on $\tau$; taking $\tau$ derivatives to
both sides of the equation we get
\begin{equation}
    \sfrac{\mathrm{d}}{\mathrm{d}\tau} \left(\sfrac{\dot{r}}{r} \right)^2 \approx
    2 \mu \Omega_c \sfrac{\dot{\tau}^2}{\tau^3}.
\end{equation}
We conclude that the expansion rate has a positive derivative,
indicating that the Universe expands faster with the passage of
time. This effect has been observed and is usually attributed to a
cosmological constant, however we see here that it can be explained
geometrically, if we allow for a small mass density $\Omega = \mu
\Omega_c$.

Tests are needed to confirm that the recent supernovae observations
are compatible with the predictions from the equation above. The
latter can be simplified if we make some further approximations. We
can replace $\dot{r}/r$ and $1/\tau$ by the Hubble parameter $H$,
while making $\dot{\tau} \approx 1$; we can also take the $\tau$
derivative to be approximately a derivative against $t$. With those
approximations the new equation is
\begin{equation}
    \dot{H} \approx 2 \mu \Omega_c H^2.
\end{equation}
All parameters in this equation are known from observation, with the
exception of $\mu$, which is the mass density expressed as a
fraction of $\Omega_c$; this is expected to be around $5\%$.

\section{Conclusion}
The search for a unified formulation of physics has led the author
to explore 5-dimensional spacetime endowed with curvature; this
space is best described by the associated geometric algebra
$G_{4,1}$, which the author uses for the study of monogenic
functions. These are functions that zero the fundamental vector
derivative; as such they play a fundamental role in the geometry.
Monogenic functions in 5-dimensional spacetime have two important
consequences: they induce an hyperspherical symmetry and they model
dynamics.

The identification of an overall hyperspherical symmetry in the
Universe is all that is needed to derive the Hubble relation, which
then appears as a purely geometrical effect. The mysterious critical
density, assigned to dark matter in the standard model of cosmology,
is here given a geometrical explanation. Monogenic functions can be
applied in curved geometries and show the ability to model
electrodynamics and relativistic dynamics. Assuming that a mass
distribution induces space curvature, we are able to show that
orbits with flat rotation curves, such as are found in most
galaxies, become possible; the author contends that galaxies can be
modelled by a single monogenic function, avoiding the recourse to
dark matter. Assuming the whole Universe to be governed by one
monogenic function, it is shown that a small mass density induces an
accelerated expansion, which may avoid the embarrassing appeal to a
cosmological constant.


\bibliographystyle{unsrtbda}
\bibliography{Abrev,aberrations,assistentes}

\end{document}